\magnification=1200
\def\lbrk{\hfil\break}
\countdef\pageno=0

\def\sspacing{\baselineskip=.21875truein \lineskip=.04truein
\lineskiplimit=.03truein}
\def\hspacing{\baselineskip=.265truein \lineskip=.0775truein
\lineskiplimit=.07truein}

\def\author#1{\par\vskip12pt\par\centerline{#1}}

\def\xskip{\hbox to .125truein{}}

\def\half{{\textstyle {1\over 2}}}

\def\fourth{{\textstyle {1\over 4}}}
\def\TEX{\hbox{\:aT\hskip-2pt\lower1.94pt\hbox{E}\hskip-2pt X}}
\def\gsim{\mathrel{\rlap{\raise 1.5pt \hbox{$>$}}\lower 3.5pt
\hbox{$\sim$}}}
\def\lsim{\mathrel{\rlap{\raise 2.5pt \hbox{$<$}}\lower 2.5pt
\hbox{$\sim$}}}
 
\def\boxit#1{\vbox{\hrule\hbox{\vrule\hskip 5pt
\def\hb{\rlap{\raise 5.5 pt
\hbox{\vrule height 0.2pt width 4pt depth 0pt}}h}
\vbox{\vskip 5pt #1 \vskip 5pt}\hskip 5pt\vrule}\hrule}}

\def\Re{{\rm Re\ }}
 
\def\simlim#1#2{\rlap{\lower 6pt
\hbox{$\null_{#1}$}}{\hbox to #2truein{\hfil
$\sim$\hfil}}}
\def\simeqlim#1#2{\rlap{\lower 6pt
\hbox{$\null_{#1}$}}{\hbox to #2truein{\hfil
$\simeq$\hfil}}}
 
\voffset=1.0truecm
\hsize=16.0truecm
\hoffset=0.5truecm
%
 
\def\nolabels{\def\eqnlabel##1{}\def\eqlabel##1{}\def\reflabel##1{}}
\def\writelabels{\def\eqnlabel##1{%
{\escapechar=` \hfill\rlap{\hskip.09in\string##1}}}%
\def\eqlabel##1{{\escapechar=` \rlap{\hskip.09in\string##1}}}%
\def\reflabel##1{\noexpand\llap{\string\string\string##1\hskip.31in}}}
\nolabels
%
\global\newcount\secno \global\secno=0
\global\newcount\subsecno \global\subsecno=0
\global\newcount\meqno \global\meqno=1
\font\fourteenbf=cmbx10 scaled \magstep2
 
\def\newsec#1{\global\advance\secno by1
\xdef\secsym{\the\secno.}\global\meqno=1\global\subsecno=0
\bigbreak\bigskip 
\centerline{{\fourteenbf\the\secno. #1}}\par\nobreak\medskip\nobreak}
\xdef\secsym{}
\def\eqnref#1{\xdef #1{(\secsym\the\meqno)}\global\advance\meqno by1%
\eqno#1\eqlabel#1}
\def\eqalref#1{\xdef #1{(\secsym\the\meqno)}\global\advance\meqno by1%
&#1\eqlabel#1}
 
 
\global\newcount\refno \global\refno=1
\newwrite\rfile
 
\def\ref#1#2{[\the\refno]\nref#1{#2}}
\def\nref#1#2{\xdef#1{[\the\refno]}%
\ifnum\refno=1\immediate\openout\rfile=refs.tmp\fi%
\immediate\write\rfile{\noexpand\item{#1\ }\reflabel{#1}#2}%
\global\advance\refno by1}
\def\addref#1{\immediate\write\rfile{\noexpand\item{}#1}}

\def\newsubsec#1{\par\vskip12pt plus5pt minus 3pt\goodbreak\par
\global\advance\subsecno by1
{\bf\noindent\number\secno.\number\subsecno\em #1}
\par\nobreak\medskip\nobreak}
\def\nr{\secsym\the\meqno    
\global\advance\meqno by 1}
\def\abstract#1{\par\vskip20pt\par\centerline{ABSTRACT}\par #1}

%
%
%
%
 
\catcode`@=11
\def\chkspace{%
  \relax   
  \begingroup\ifhmode\aftergroup\dochksp@ce\fi\endgroup}
\def\dochksp@ce{%
  \unskip              
  \futurelet\chkspct@k\d@chkspc  
}
\def\d@chkspc{%
  \let\nxtsp@ce=\relax
  \ifx\chkspct@k.\else     
    \ifx\chkspct@k,\else
      \ifx\chkspct@k;\else
        \ifx\chkspct@k!\else
          \ifx\chkspct@k?\else
            \ifx\chkspct@k:\else
              \ifx\chkspct@k)\else
              \ifx\chkspct@k(\else
                \ifx\chkspct@k]\else
                  \ifx\chkspct@k-\else
                    \ifx\chkspct@k\egroup\else  
                      \let\nxtsp@ce=\put@space  
                    \fi
                  \fi
                \fi
              \fi
              \fi
            \fi
          \fi
        \fi
      \fi
    \fi
  \fi
  \nxtsp@ce
}
\def\put@space{$\;$}
\catcode`@=12

\def\ra{{$\rightarrow$}\chkspace}
\def\etal{{\it et al.}\chkspace}

\def\adhoc{{\it ad hoc}\chkspace}
\def\ie{{\it i.e.}\chkspace}

\def\eg{{\it eg.}\chkspace}

\def\ibid{{\it ibid}\chkspace}

\def\apriori{{\it a priori}\chkspace}

\def\ep{{e$^+$e$^-$}\chkspace}
\def\epa{{e$^+$e$^-$ annihilation}\chkspace}

\def\qu{\quad}

\def\gluino{\relax\ifmmode \tilde{g} \else $\tilde{g}$ \fi\chkspace}
 
\def\qq{q$\overline{\rm q}$\chkspace}

\def\bb{\relax\ifmmode {\rm b}\bar{\rm b}
       \else ${\rm b}\bar{\rm b}$ \fi\chkspace}
\def\cc{\relax\ifmmode {\rm c}\bar{\rm c}
       \else ${\rm c}\bar{\rm c}$ \fi\chkspace}
\def\tt{\relax\ifmmode {\rm t}\bar{\rm t}
       \else ${\rm t}\bar{\rm t}$ \fi\chkspace}

\def\qqg{\relax\ifmmode {\rm q}\overline{\rm q}{\rm g}
\else q$\overline{\rm q}$g \fi\chkspace}

\def\afb{\relax\ifmmode A_{FB} \else
{{$A_{FB}$}}\fi\chkspace}
\def\afbb{\relax\ifmmode A_{FB}^b \else
{{$A_{FB}^b$}}\fi\chkspace}
\def\pafb{\relax\ifmmode \tilde{A}_{FB} \else
{{$\tilde{A}_{FB}$}}\fi\chkspace}
\def\pafbb{\relax\ifmmode \tilde{A}_{FB}^b \else
{{$\tilde{A}_{FB}^b$}}\fi\chkspace}
 
\def\pafbzo{\relax\ifmmode \tilde{A}_{FB}|_{O(0)} \else
{{$\tilde{A}_{FB}|_{O(0)}$}}\fi\chkspace}
\def\pafbfo{\relax\ifmmode \tilde{A}_{FB}|_{\oalp} \else
{{$\tilde{A}_{FB}|_{\oalp}$}}\fi\chkspace}
\def\pafbso{\relax\ifmmode \tilde{A}_{FB}|_{\oalpsq} \else
{{$\tilde{A}_{FB}|_{\oalpsq}$}}\fi\chkspace}
\def\pafbto{\relax\ifmmode \tilde{A}_{FB}|_{\oalpc} \else
{{$\tilde{A}_{FB}|_{\oalpc}$}}\fi\chkspace}
 
\def\pafbbzo{\relax\ifmmode \tilde{A}_{FB}^b|_{O(0)} \else
{{$\tilde{A}_{FB}^b|_{O(0)}$}}\fi\chkspace}
\def\pafbbfo{\relax\ifmmode \tilde{A}_{FB}^b|_{\oalp} \else
{{$\tilde{A}_{FB}^b|_{\oalp}$}}\fi\chkspace}
\def\pafbbso{\relax\ifmmode \tilde{A}_{FB}^b|_{\oalpsq} \else
{{$\tilde{A}_{FB}^b|_{\oalpsq}$}}\fi\chkspace}
\def\pafbbto{\relax\ifmmode \tilde{A}_{FB}^b|_{\oalpc} \else
{{$\tilde{A}_{FB}^b|_{\oalpc}$}}\fi\chkspace}
 
\def\afbo0{\tilde{A}_{FB}|_{O(0)}}
\def\afbo1{\tilde{A}_{FB}|_{\oalp}}
\def\afbo2{\tilde{A}_{FB}|_{\oalpsq}}
\def\afbo3{\tilde{A}_{FB}|_{\oalpc}}
 
\def\lam{\relax\ifmmode \Lambda_{\overline{MS}}
       \else {{$\Lambda_{\overline{MS}}$}}\fi\chkspace}
\def\lamuds{\relax\ifmmode \Lambda^{(3)}_{\overline{MS}}
       \else {{$\Lambda^{(3)}_{\overline{MS}}$}}\fi\chkspace}
\def\lamudsc{\relax\ifmmode \Lambda^{(4)}_{\overline{MS}}
       \else $\Lambda^{(4)}_{\overline{MS}}$\fi\chkspace}
\def\lamudscb{\relax\ifmmode \Lambda^{(5)}_{\overline{MS}}
       \else $\Lambda^{(5)}_{\overline{MS}}$\fi\chkspace}

\def\alp{\relax\ifmmode \alpha_s\else $\alpha_s$\fi\chkspace}
\def\alpbar{\relax\ifmmode \overline{\alpha_s}
       \else $\overline{\alpha_s}$\fi\chkspace}
\def\alpmz{\relax\ifmmode \alpha_s(M_Z)\else $\alpha_s(M_Z)$\fi\chkspace}
\def\alpmzsq{\relax\ifmmode \alpha_s(M_Z^2)
       \else $\alpha_s(M_Z^2)$\fi\chkspace}
 
\def\oalp{\relax\ifmmode O(\alpha_s)\else{{O($\alpha_s$)}}\fi\chkspace}
\def\oalpsq{\relax\ifmmode O(\alpha_s^2)
           \else{{O($\alpha_s^2$)}}\fi\chkspace}
\def\oalpc{\relax\ifmmode O(\alpha_s^3)
           \else{{O($\alpha_s^3$)}}\fi\chkspace}
\def\oalpf{\relax\ifmmode O(\alpha_s^4)
           \else{{O($\alpha_s^4$)}}\fi\chkspace}

\def\z0{{$Z^0$}\chkspace}
\def\Dst{\relax\ifmmode {\rm D}^* \else {D$^*$}\fi\chkspace}
\def\Dpl{\relax\ifmmode {\rm D}^+ \else {D$^+$}\fi\chkspace}
\def\D0{\relax\ifmmode {\rm D}^0 \else {D$^0$}\fi\chkspace}
\def\Kst{\relax\ifmmode {\rm K}^* \else {K$^*$}\fi\chkspace}
\def\K0{\relax\ifmmode {\rm K}^0_s \else {K$^0_s$}\fi\chkspace}
\def\Kpl{\relax\ifmmode {\rm K}^+ \else {K$^+$}\fi\chkspace}
\def\Kstz{\relax\ifmmode {\rm K}^{*0} \else {K$^{*0}$}\fi\chkspace}

\input epsf
 
 
\def\qq{q$\bar{\rm q}\;$}

\newcount\notenumber

\def\note{\advance\notenumber by 1
  \footnote{$^{\#{\the\notenumber}}$}}
 
\def\fig#1#2{\sspacing\hangindent=.75truein
\noindent \hbox to .75truein{Fig.\ #1.\hfil}#2
\hspacing\vskip 10pt}
 
\def\eqqqalignno#1{\tabskip=0em plus 50em\openup2\jot
\halign to \hsize{\hfil$
\displaystyle{##\null}$\tabskip=0pt &$
\displaystyle{##\qquad}$\hfil\tabskip=0pt &\hfil$
\displaystyle{##\null}$\tabskip=0pt &$\displaystyle{##}$\hfil\tabskip=1em
plus 50em &\hfil{##}\tabskip=0pt \crcr #1\crcr}}
 
\def\d{{\rm d}}

\def\eg{{\it e.g.}} 
 
\def \sup{^{\vphantom{2}}}
 
\def\sigmaU{\sigma\sup_{\rm U}}
\def\sigmaL{\sigma\sup_{\rm L}}
\def\sigmaT{\sigma\sup_{\rm T}}
\def\sigmaI{\sigma\sup_{\rm I}}
\def\sigmaA{\sigma\sup_{\rm A}}
\def\sigmaP{\sigma\sup_{\rm P}}
 
\def\hatsigmaU{\hat\sigma\sup_{\rm U}}
\def\hatsigmaL{\hat\sigma\sup_{\rm L}}
\def\hatsigmaT{\hat\sigma\sup_{\rm T}}
\def\hatsigmaI{\hat\sigma\sup_{\rm I}}
\def\hatsigmaA{\hat\sigma\sup_{\rm A}}
\def\hatsigmaP{\hat\sigma\sup_{\rm P}}

\def\tilAFB{\tilde A_{\rm FB}}

\def\FU{F\sup_{\rm U}}
\def\FL{F\sup_{\rm L}}
\def\FT{F\sup_{\rm T}}
\def\FI{F\sup_{\rm I}}
\def\FA{F\sup_{\rm A}}
\def\FP{F\sup_{\rm P}}

\def\thW{\theta_W}
 
\hspacing

{\baselineskip=0pt
\hfill{SLAC--PUB--7401}
 
\hfill{MIT-LNS-97-237}
 
\hfill{January 1997}}
 
\vskip .6truecm
 
\centerline{\bf THREE-JET EVENT ORIENTATION}
\centerline{\bf IN \ep ANNIHILATION:}
\centerline{\bf NEW TESTS OF THE STANDARD MODEL$^{\star}$}
 
\vskip 1truecm
 
\centerline{\bf P.N. Burrows$^{**}$}
 
\vskip .3truecm
 
\centerline{\it Stanford Linear Accelerator Center}
\centerline{\it Stanford University, Stanford, CA94309, USA}
 
\vskip .2truecm
 
\centerline{burrows@slac.stanford.edu}

\vskip .4truecm
 
\centerline{\bf P. Osland}
 
\vskip .3truecm
 
\centerline{\it Department of Physics, University of Bergen}
\centerline{\it All\'egt.~55, N-5007 Bergen, Norway}
 
\vskip .2truecm
 
\centerline{per.osland@fi.uib.no}
 
\vskip .5truecm
 
\centerline{\bf Abstract}
 
{2}      
 
{\baselineskip=4pt
 
\noindent
We discuss the orientation of \ep \ra \qqg events
in terms of the polar and azimuthal angles of the
event plane w.r.t.\ the electron beam direction.
We define an asymmetry of the azimuthal-angle distribution
which, along with the left-right forward-backward
polar-angle asymmetry, is sensitive to parity-violating
effects in three-jet events;
these have yet to be explored experimentally.
We have evaluated these observables at \oalp in perturbative
QCD and present their dependence on longitudinal beam polarisation
and c.m.\ energy.
We also define a moments analysis in terms of the orientation
angles that allows a new and
more detailed test of QCD by isolating the six
independent helicity cross-sections.
 
\vskip .5truecm
 
\centerline{\it Submitted to Physics Letters B}
 
\vfil
 
\noindent
$\star$ {Work supported by Department of Energy contracts
DE--FC02--94ER40818 (MIT)
and   DE--AC03--76SF00515 (SLAC).}
 
\noindent
$**$ {Permanent address: Lab. for Nuclear Science,
M.I.T., Cambridge, MA 02139, USA.}
 
}
 
\eject
 
\noindent{\bf 1. Introduction}
 
\vskip 1\baselineskip
 
\noindent
In \ep annihilation, events containing three distinct jets of hadrons
were first observed many years ago at the PETRA storage ring
\ref{\refPETRA}
{TASSO Collab., R. Brandelik \etal, Phys. Lett. {\bf 86B}
(1979) 243.\hfill\break
Mark J Collab., D.P. Barber \etal, Phys. Rev. Lett. {\bf 43}
(1979) 830.}
\addref
{PLUTO Collab., Ch. Berger \etal, Phys. Lett. {\bf 86B}
(1979) 418.\hfill\break
JADE Collab., W. Bartel \etal, Phys. Lett. {\bf 91B}
(1980) 142.}.
Such events were interpreted in terms of the fundamental
process \ep \ra \qqg\ (Fig.~1),
providing direct evidence for the existence of the gluon,
the vector boson of the theory of strong interactions, Quantum
Chromodynamics (QCD)
\ref{\refQCD}
{H. Fritzsch, M. Gell-Mann and H. Leutwyler, Phys.\ Lett.\
{\bf 47B} (1973) 365; \lbrk
D.J.\ Gross and F. Wilczek, Phys.\ Rev.\ Lett.\ {\bf 30} (1973) 1343;
\lbrk
H.D. Politzer, Phys.\ Rev.\ Lett.\ {\bf 30} (1973) 1346; \lbrk
S. Weinberg, Phys.\ Rev.\ Lett.\ {\bf 31} (1973) 494.}.
A large number of subsequent studies of the properties
of such events
\ref{\refWU}
{For a review see S.L.~Wu, Phys. Rep. {\bf 107} (1984) 59.}\
has verified this interpretation.
 
Here we consider the orientation of the \qqg plane or `event plane'
in terms of the angles $\theta$ and $\chi$ (Fig.~2),
where $\theta$ is the polar angle of
the quark direction with respect to the electron beam, and $\chi$ is
the azimuthal orientation angle of the event plane with respect to the
quark-electron plane, such that:
$$
\cos\chi\qu = \qu {\vec q \times \vec g \over{
| \vec q \times \vec g |}}
\cdot {\vec q \times \vec{e^-} \over {
| \vec q \times \vec{e^-}|}}.
\eqnref{\eqchi}
$$
The polar angle can also be defined in two-jet
events of the type \ep \ra \qq, in which case the distribution
in $\theta$ is determined in the electroweak theory
\ref{\refEW}
{See, \eg, G. Altarelli \etal, in
Physics at LEP, CERN 86-02 (1986),
eds.\ J. Ellis and R. Peccei, p.~1.},
and displays a c.m.\ energy-dependent
forward-backward asymmetry which has been observed in many
experiments
\ref{\refAFBREV}
{See, \eg, T. Kamae,
in Proc.\ XXIV International Conference on High Energy Physics,
Munich, August 4--10, 1988 (eds.\ R. Kotthaus and J.H. K\"uhn,
Springer Verlag, 1989)
p.~156.},
\ref{\refBLONDEL}
{
A. Blondel, to appear in Proc. XXVIII International
Conference on High Energy Physics, Warsaw, Poland, July 25-31 1996.
}.
For \ep annihilation at the \z0 resonance, the polar-angle asymmetry
is large only if one, or both, of the beams is longitudinally
polarised, as at SLC/SLD
\ref{\refSLDAFB}
{SLD Collab., K. Abe \etal, Phys.\ Rev.\ Lett.\ {\bf 74} (1995) 2890;
{\it ibid} {\bf 74} (1995) 2895;
{\it ibid} {\bf 75} (1995) 3609.}.
The azimuthal angle $\chi$ is, of course, undefined in \qq events,
but in \qqg events, as we shall show, it also displays an
asymmetry which can be large at the \z0 resonance in
the case of highly polarised electrons.
This azimuthal-angle distribution has not yet been investigated
experimentally.
 
We review (Section 2) the fully-differential cross-section for
three-jet production in \ep annihilation, and present (Section 3)
the polar- and azimuthal-angle distributions, illustrating
their dependence on the longitudinal electron beam polarisation
and the c.m.\ energy.
In Section 4 we consider asymmetries of these angular distributions,
which provide a currently unexplored search ground for anomalous
parity-violating effects in \qqg events.
In Section 5 we define moments of the cross-section in terms of
$\cos\theta$ and $\cos\chi$, which would allow one to make
a more detailed test of QCD by determining the six independent
helicity cross-sections, which have not yet been directly
explored experimentally.
In Section 6 we discuss more inclusive cases in which the requirement
of quark and antiquark jet identification is relaxed,
and in Section 7 we present a summary and concluding remarks.

\vskip 1\baselineskip
 
\noindent{\bf 2. Review of the \ep \ra \qqg Differential Cross-Section}
 
\vskip .7\baselineskip
 
\noindent
Firstly, we give a brief review of the differential
cross-section for three-jet production in \epa
at c.m.\ energy $\sqrt{s}$, assuming massless partons.
Let $\vec q$, $\vec {\bar q}$, and $\vec g$ denote the quark, antiquark
and gluon momenta respectively, and
$x$, $\bar x$ and $x_g$ be the scaled energies:
$$
|\vec{q}\,|=x\sqrt{s}, \qquad |\vec{\bar q}\,|=\bar x\sqrt{s}, \qquad
|\vec{g}\,|=x_g\sqrt{s},
\eqnref{\eqxxx}
$$
with  $x+\bar x+x_g=2.$
Allowing for longitudinal beam polarisation\note{Expressions for
the general case of arbitrarily polarised beams are given in
\ref{\refOOO}
{H.A. Olsen, P. Osland, I. \O verb\o,
Nucl. Phys. {\bf B171} (1980) 209.}.},
the fully-differential three-jet cross-section can then be expressed as
{\refOOO}:
$$\eqalignno{
2\pi{\d^4\sigma \over \d(\cos\theta)\d\chi\d x\d\bar x}
&= \biggl[
{3\over8}(1+\cos^2\theta)\, {\d^2\sigmaU\over \d x\d\bar x}
+{3\over4}\sin^2\theta\, {\d^2\sigmaL\over \d x\d\bar x} \cr
&+{3\over4}\sin^2\theta\cos2\chi\, {\d^2\sigmaT\over \d x\d\bar x}
+{3\over2\sqrt{2}}\sin2\theta\cos\chi\, {\d^2\sigmaI\over \d x\d\bar x}
\biggr] h_f^{(1)}(s) \cr
&+ \biggl[
{3\over4}\cos\theta\, {\d^2\sigmaP\over \d x\d\bar x}\;
-\;{3\over\sqrt{2}}\sin\theta\cos\chi\, {\d^2\sigmaA\over \d x\d\bar x}
\biggr]h_f^{(2)}(s),
\eqalref{\eqfour}}$$
where at lowest order in the electroweak theory
the dependences on flavour and beam polarisation
are given by the functions:
$$\eqalignno{
h_f^{(1)}(s) &=
Q_f^2\Xi -2Q_f\Re f(s)(v\Xi-a\xi)v_f \cr
&+|f(s)|^2[(v^2+a^2)\Xi-2va\xi](v_f^2+a_f^2), \eqalref{\eqhfone}\cr
h_f^{(2)}(s) &=
-2Q_f\Re f(s)(a\Xi-v\xi)a_f \cr
&+|f(s)|^2[-(v^2+a^2)\xi+2va\Xi]2v_fa_f, \eqalref{\eqhftwo}}$$
with
$$
f(s)={1\over 4\sin^22\thW}\; {s\over s-M_Z^2+iM_Z\Gamma_Z^{\rm tot}},
\eqno(\nr)
$$
where $Q_f$ is the charge of quark flavour $f$;
$v$, $a$ ($v_f$, $a_f$) are the vector and axial vector couplings of
the \z0 to the electron (quark of flavour $f$), respectively,
$$
\Xi=1-P_-^{\|}P_+^{\|}, \qquad \xi=P_-^{\|}-P_+^{\|}, \eqno(\nr)
$$
and $P_-^{\|}$ ($P_+^{\|}$) is  the longitudinal polarisation
of the electron (positron) beam.
 
Several important points concerning eq.~{\eqfour} should be noted.
The cross-section can be written
as a sum of 6 terms, each of which may be factorised
into three contributions: the first factor
is a simple trigonometric function
of the polar and azimuthal orientation angles
$\theta$ and $\chi$, and the second,
$\d^2\sigma_i/\d x \d \bar x$ ($i$=U, L, T, I, P, A),
is a function of the parton momentum fractions;
these are determined by QCD and kinematics;
the third factor, $h_f^{(1,2)}(s)$,
is a function containing the dependence on the fermion electroweak
couplings.
Hence, in each term there is factorisation both between
the dynamical contributions of the QCD and electroweak sectors of the
Standard Model and between the orientation of the event plane and the
relative orientation of the jets within the plane.
We exploit this
property later by defining moments in terms of
$\cos\theta$ and $\cos\chi$ in order to isolate the different terms.
The $\sigma_i$ are often referred to in the literature
as {\it helicity cross-sections}\note{Helicity cross sections
have also been discussed in the context of single-particle
inclusive fragmentation functions
\ref{\refNaWe}
{P. Nason, B.R. Webber, Nucl.\ Phys.\ {\bf B421} (1994) 473;
Erratum: \ibid\ {\bf B480} (1996) 755.}.},
and the form of eq.~{\eqfour}, with
six terms, each containing one of the six
independent helicity cross-sections, has been shown
\ref{\refKS}
{J.G. K\"orner, G.A. Schuler, Z. Phys.\ {\bf C26} (1985) 559.}\
to be valid for massless partons up to \oalpsq in perturbative QCD.
 
At \oalp in perturbative QCD
{\refOOO},
\ref{\refLaermann}
{E. Laermann, K.H. Streng, P.M. Zerwas,
Z. Phys.\ {\bf C3} (1980) 289;
Erratum: Z. Phys.\ {\bf C52} (1991) 352.}:
$$
{\d^2\sigma_i\over\d x \d \bar x} \quad=\quad
{\displaystyle{\tilde\sigma\over(1-x)(1-\bar x)}}\, F_i,
\eqnref{\eqthree}
$$
where
$$
\tilde\sigma={4\pi\alpha^2\alpha_s(s)\over3\pi s}.
\eqno(\nr)
$$
For the coordinate system of Fig.~2 one has {\refOOO}:
$$\eqqqalignno{
\FU &=2 x^2 +\bar x^2(1+\bar c^2), &
%
\FL &=\bar x^2{\bar s}^2, \cr
\FT &=\half \FL, &
%
\FI &=2^{-1/2} \bar x^2\bar c\bar s, \cr
%
\FA &=2^{-1/2} \bar x^2\bar s, &
%
\FP &=2(x^2-\bar x^2\bar c), \eqalref{\eqeleven}}
$$
where
$\overline c \, =\,\cos(\psi_{q\bar{q}})$ and
$\overline s \, =\, \sin(\psi_{q\bar{q}})$, with
$\psi_{q\bar{q}}$ the angle between the quark and antiquark
momenta. Since the quark and antiquark tend to have a small
acollinearity angle, $\FL$, $\FT$, $\FI$ and $\FA$ will typically
be small compared with $\FU$ and $\FP$.

\vfill\eject
\vskip 1\baselineskip
 
\noindent{\bf 3. Polar- and Azimuthal-Angle Distributions}
 
\vskip .7\baselineskip
 
\noindent
We now discuss the singly-differential cross-sections in terms of
$\cos\theta$ or $\chi$.
Consider integrating eq.~{\eqfour} first over $x$ and
$\bar{x}$, with the integration domain given by a standard
jet resolution criterion $y_c$\note{Any of the six infra-red-
and collinear-safe jet-definition algorithms `E', `E0', `P',
`P0', `D' and `G'
\ref{\refSB}
{S. Bethke \etal, Nucl.\ Phys.\ {\bf B370} (1992) 310.}\
could be used. We have used the `E' procedure which,
at \oalp, gives identical results to the `E0', `P' and `P0'
procedures.}
and using
the notation:
$$
\hat\sigma_i \qu \equiv \qu
\int_{y_c}\d x\int\d\bar x
{\d^2\sigma_i \over \d x\d\bar x}. \eqnref{\eqycut}$$
Integrating over $\chi$ we then obtain:
$$     {\d  \sigma \over \d(\cos\theta)}
 = \left( {3\over8}(1+\cos^2\theta)\, \hat \sigmaU
+{3\over4}\sin^2\theta\,  \hat\sigmaL \right) h_f^{(1)}(s)
 +{3\over4}\cos\theta\,  \hat\sigmaP \, h_f^{(2)}(s),
\eqnref{\eqathe}
$$
where the term containing $\hat\sigmaP$ represents the well-known quark
forward-backward asymmetry resulting from parity violation in the weak
interaction, but for the three-jet case.
Similarly, by integrating over $\cos\theta$ we obtain:
$$
2\pi\, {\d \sigma \over \d\chi}
= \left( \hat \sigmaU + \hat \sigmaL
+\cos 2\chi\, \hat \sigmaT \right) h_f^{(1)}(s)
-{3\pi\over2\sqrt{2}}\, \cos\chi\, \hat\sigmaA \, h_f^{(2)}(s),
\eqnref{\eqachi}
$$
where the term containing $\hat\sigmaA$ represents an azimuthal,
parity-odd asymmetry analogous to the last term in
eq.~{\eqathe} but owing its existence to the radiation of the gluon.
 
For the case of longitudinally-polarised electrons and unpolarised
positrons the dependences of these singly-differential distributions
on the beam polarisation and c.m.\
energy are illustrated in Figs.~3(a,b) and 4(a,b) respectively.
We calculated the $\sigma_i$ at \oalp as in Ref.~{\refOOO}.
Fig.~3a shows the distribution in
$\cos\theta$ at $\sqrt{s}$ = $M_Z$ for down-type
quarks, $y_c=0.02$ and electron
longitudinal polarisation $p$ = $+1$, 0 and $-1$.\note{We refer to
positive (negative) polarisation as right- (left-) handed respectively.}
The current SLC/SLD case of $p$ = $\pm0.77$
\ref{\refSLDninetysix}
{
SLD Collab., K. Abe \etal, SLAC-PUB-7291 (1996);
subm.\ to Phys.\ Rev.\ Lett.
}\
is also indicated. The quark polar-angle forward-backward
asymmetry is large for high beam polarisation, and its sign
changes with the sign of the polarisation.
The less familiar azimuthal-angle distribution is shown
in Fig.~3b for the same cases as in Fig.~3a;
the distribution is symmetric about $\chi=\pi$.
The phase change of the
$\chi$ distribution when the beam polarisation sign is changed is a
reflection of the sign reversal of the forward-backward asymmetry in
$\cos\theta$. Qualitatively similar results are
obtained for up-type quarks, and for other values of $y_c$.
 
We illustrate the energy dependence of the $\cos\theta$- and
$\chi$-distributions in Figs.~4a and 4b, respectively,
for down-type quarks at fixed electron polarisation $p$ = $+1$,
with results at $\sqrt{s}$ = 35, 60, 91 and 200 GeV,
corresponding to \epa at the PETRA, TRISTAN, SLC/LEP and
LEP2 collider energies.
The variation with energy is due to the varying relative
contribution of $\gamma$ and \z0 exchange in the \epa process.
Results are also shown for a possible high-energy collider operating
with polarised electrons at $\sqrt{s}$ = 500 GeV and 2 TeV.
If such a facility could be operated at lower energies, where,
apart from $\sqrt{s}=91$~GeV (SLC), polarised beams were
not previously available, measurements in the same experiment
of the distributions shown in Figs.~4(a,b) would
provide a significant consistency check of the Standard Model.

\vskip 1\baselineskip
 
\noindent{\bf 4. Polar- and Azimuthal-Angle Asymmetries}
 
\vskip .7\baselineskip
 
\noindent
By analogy with the {\it left-right forward-backward asymmetry}
of the polar-angle distribution:
 
$$\eqalignno{
&\tilAFB(|p|)|_{\cos\theta}  \eqalref{\eqAFBtheta} \cr
&\equiv
{
\int_0^1\,{\d\sigma^{\rm L}\over \d\cos\theta}\,\d\cos\theta
-
\int_{-1}^0\,
          {\d\sigma^{\rm L}\over \d\cos\theta}\,\d\cos\theta
-\left(
\int_0^1\,{\d\sigma^{\rm R}\over \d\cos\theta}\,\d\cos\theta
-
\int_{-1}^0\,
          {\d\sigma^{\rm R}\over \d\cos\theta}\,\d\cos\theta
\right)
\over{
\int_0^1\,{\d\sigma^{\rm L}\over \d\cos\theta}\,\d\cos\theta
+
\int_{-1}^0\,
          {\d\sigma^{\rm L}\over \d\cos\theta}\,\d\cos\theta
+
\int_0^1\,{\d\sigma^{\rm R}\over \d\cos\theta}\,\d\cos\theta
+
\int_{-1}^0\,
          {\d\sigma^{\rm R}\over \d\cos\theta}\,\d\cos\theta
}},
\cr
}
$$
we define a corresponding asymmetry of the azimuthal-angle
distribution:
$$
\tilde A(|p|)|_{\chi}\;
\equiv\;
{
\int_{{\pi\over 2}}^{\pi}
\,{\d\sigma^{\rm L}\over \d\chi}\,\d\chi
-
\int_0^{{\pi\over 2}}
\,{\d\sigma^{\rm L}\over \d\chi}\,\d\chi
-\left(
\int_{{\pi\over 2}}^{\pi}
\,{\d\sigma^{\rm R}\over \d\chi}\,\d\chi
-
\int_0^{{\pi\over 2}}
          {\d\sigma^{\rm R}\over \d\chi}\,\d\chi
\right)
\over{
\int_{{\pi\over 2}}^{\pi}
\,{\d\sigma^{\rm L}\over \d\chi}\,\d\chi
+
\int_0^{{\pi\over 2}}
\,{\d\sigma^{\rm L}\over \d\chi}\,\d\chi
+
\int_{{\pi\over 2}}^{\pi}
\,{\d\sigma^{\rm R}\over \d\chi}\,\d\chi
+
\int_0^{{\pi\over 2}}
\,{\d\sigma^{\rm R}\over \d\chi}\,\d\chi
}},
\eqnref{\eqAFBchi}
$$
where $\sigma^{\rm L,R}=\sigma(\mp|p|)$ is the \ep \ra \qqg cross-section
for a left- (L) or right- (R) handed electron beam of polarisation
magnitude $|p|$.
 
For the case of \epa at the \z0 resonance using electrons of
longitudinal polarisation $p$ and unpolarised positrons,
as at SLC, eqs.~{\eqhfone} and {\eqhftwo} reduce to the simple forms
$$\eqalignno{
h_f^{(1)}(M_Z^2)  &=
 |f(M_Z^2)|^2[(v^2+a^2)   -2vap  ](v_f^2+a_f^2), \cr
h_f^{(2)}(M_Z^2)  &=
 |f(M_Z^2)|^2[-(v^2+a^2)p  +2va   ]2v_fa_f,  \eqalref{\eqsimple}
}$$
and hence
$$\eqalignno{
\tilAFB(|p|)|_{\cos\theta}\qu
&=\qu {3\over4}\;|p|\;{\hat\sigmaP\over{\hat\sigmaU+\hat\sigmaL}}\;A_f, \cr
\tilde A(|p|)|_{\chi}\qu&=\qu {3\over\sqrt{2}}\;|p|\;
{\hat\sigmaA\over{\hat\sigmaU+\hat\sigmaL}}\;A_f,
\eqalref{\eqAFBboth}\cr
}
$$
where we use the common notation $A_f\; \equiv\; 2v_f a_f/(v_f^2+a_f^2)$.
Whereas both asymmetries are directly proportional to the beam
polarisation $|p|$ and the electroweak coupling $A_f$,
the $\cos\theta$ asymmetry is proportional to the
helicity cross-section $\hat\sigmaP$, and
the $\chi$ asymmetry to the helicity cross-section $\hat\sigmaA$.
Since the electroweak factor $A_f$ is predicted to a high degree of
accuracy by the Standard Model, and, in the case of b and c quarks,
has been measured using predominantly \qq final states at SLC and LEP
{\refBLONDEL},
measurement of these asymmetries in \qqg events at
SLC/SLD would allow one to probe $\hat\sigmaP$ and $\hat\sigmaA$,
which have yet to be investigated experimentally. Furthermore,
the ratio of the asymmetries is independent of both
polarisation and electroweak couplings and depends only on the ratio
of $\hat\sigmaP$ and $\hat\sigmaA$:
$$
{\tilAFB(|p|)|_{\cos\theta}\over{\tilde A(|p|)|_{\chi}}}
\qu=\qu {\sqrt{2}\over 4}\;{\hat\sigmaP\over{\hat\sigmaA}}.
\eqnref{\eqratio}
$$
As a consequence of there being, up to \oalpsq in massless
perturbative QCD, only the six independent
helicity cross-sections given in eq.~{\eqfour},
the relations {\eqAFBboth} and {\eqratio} are valid up to the same order.
 
We have calculated at \oalp the ratios
$\hat\sigmaP/(\hat\sigmaU+\hat\sigmaL)$ and
$\hat\sigmaA/(\hat\sigmaU+\hat\sigmaL)$
and show in Fig.~5a their dependence on $y_c$; the dependence is weak.
For completeness we also show
$\hat\sigmaT/(\hat\sigmaU+\hat\sigmaL)$ and
$\hat\sigmaI/(\hat\sigmaU+\hat\sigmaL)$.
It would be worthwhile to investigate the size of higher-order
perturbative QCD contributions by evaluating these ratios at \oalpsq;
this should, in principle, be possible using the matrix elements
described in Ref.~\ref{\refCG}
{
S. Catani, M. Seymour, CERN-TH-96-029 (1996).
}.
It should then also be possible to estimate
\oalpc contributions using \adhoc theoretical procedures \ref{\refADHOC}
{
M.A. Samuel, G. Li, E. Steinfelds, Phys.\ Rev.\ {\bf D48} (1993) 869;
\hfill\break
A.L. Kataev, V.V. Starshenko, Mod.\ Phys.\ Lett.\ {\bf A10} (1995) 235.
}.
It would also be worthwhile to investigate quark mass effects; this
could be done at \oalp using an available calculation \ref{\refSO}
{
J.B.~Stav, H.A.~Olsen, Phys.\ Rev.\ {\bf D50} (1994) 6775.
},
and at \oalpsq when the corresponding matrix elements \ref{\refBBB}
{
A. Ballestrero, E. Maina, S. Moretti, Phys.\ Lett.\ {\bf B294} (1992) 425;
\hfill\break
M. Bilenkii, G. Rodrigo, A. Santamaria, Nucl.\ Phys.\ {\bf B439} (1995)
505; \hfill\break
A. Brandenburg, private communications.
}\
become available.
 
If higher-order perturbative contributions and mass corrections
were under control, one could confront the theoretical predictions with
experimental measurements. Significant deviations of the data from the
predictions for the asymmetries, eqs.~{\eqAFBboth}, would indicate
anomalous parity-violating contributions to the process \ep \ra \qqg,
and the ratio of asymmetries, eq.~{\eqratio}, being at lowest
order independent of the electroweak coupling factor $A_f$, would
help to unravel the dynamical origin of any effect. For example, an
\adhoc modification of strong interactions including an explicitly
parity-violating
quark-gluon coupling proportional to $(1-\epsilon \gamma_5)$ leads to
O($\epsilon$) corrections to
$\hat\sigma_i$ for $i$ = P and A, as well as
$i$ = U, L, T and I; it also leads to an additional
small term proportional to $\epsilon\sin\theta \sin\chi$ \ref{\refOG}
{
O.M. {\O}greid, Cand. Scient. Thesis, University of Bergen, October 1993
(unpublished).
}.
 
\vskip 1\baselineskip
 
\noindent{\bf 5. Moments of the Angular Distributions}
 
\vskip .7\baselineskip
 
\noindent
The potential sensitivity of the six independent helicity cross-sections
$\sigma_i$ to strong parity-violating effects\note{More
generally the $\sigma_i$ are,
of course, sensitive to the strong dynamics. For example, in
a scalar-gluon theory at $y_c=0.02$,
$\hat\sigmaA/(\hat\sigmaU+\hat\sigmaL)$ and
$\hat\sigmaP/(\hat\sigmaU+\hat\sigmaL)$
differ by factors of $-1.5$ and 0.12, respectively, relative to QCD.}
has led us to consider a generalised method for extracting them from
data.
We define moments of the cross-section {\eqfour} in terms of powers
of $\cos\theta$ and $\cos\chi$:
$$\Sigma_{mn}\equiv\int_{-1}^1\d(\cos\theta)\cos^m\theta
\int_0^{2\pi}\d\chi\cos^n\chi
\int_{y_c}\d x\int\d\bar x
{\d^4\sigma \over \d(\cos\theta)\d\chi\d x\d\bar x}.
\eqnref{\eqmoments}$$
The lowest-rank moments are given in Table~1,
where the $\hat\sigma_i$ are defined by eq.~{\eqycut}.
It should be noted that $\Sigma_{00}$ corresponds to the total
integrated cross-section for 3-jet
production, and that
$\Sigma_{01}$ and $\Sigma_{10}$ are closely related to the asymmetries
defined by eqs.~{\eqAFBtheta} and {\eqAFBchi}.
It then follows trivially that each $\hat\sigma_i$ could be
obtained from the measured moments:
$$\eqqqalignno{
\hatsigmaU \qu &= \qu \left(5\Sigma_{20}\;-\;\Sigma_{00}\right)
/h_f^{(1)}(s), &
\hatsigmaL \qu &= \qu \left(2\Sigma_{00}\;-\;5\Sigma_{20}\right)
/h_f^{(1)}(s), \cr
\hatsigmaT \qu &= \qu \left(4\Sigma_{02}\;-\;2\Sigma_{00}\right)
/h_f^{(1)}(s), &
\hatsigmaI \qu &= \qu \left({16\sqrt{2}\over 3\pi} \Sigma_{11}\right)
/h_f^{(1)}(s), \cr
\hatsigmaA \qu &= \qu \left({-4\sqrt{2}\over 3\pi} \Sigma_{01}\right)
/h_f^{(2)}(s), &
\hatsigmaP \qu &= \qu 2\Sigma_{10}/h_f^{(2)}(s),
&(\nr)}
$$
where the electroweak parameters entering via eqs.~{\eqhfone}
and {\eqhftwo} could be taken from the Standard Model,
or from measured values based predominantly on 2-jet final states.
 
\vskip 1\baselineskip
 
\noindent{\bf 6. Inclusive Cross-Sections}
 
\vskip .7\baselineskip
 
\noindent
All of the preceeding discussions have been based on the assumption
that the parton-type originator of jets is known, \ie that in \ep
\ra 3-jet events one
can identify which jet originated from the quark, antiquark and
gluon.
The definition of $\cos\theta$ requires that
the quark jet be known, whereas the definition of $\chi$
requires that both the quark jet and a second jet
origin be known. It is
difficult from an experimental point-of-view to make such
exclusive
identification for jets of hadrons measured in a detector.
Quark and antiquark jets have been identified in predominantly 2-jet
events in \ep annihilation (see \eg\ {\refSLDAFB}),
where typically only one
jet per event is tagged, with low efficiency. Identification of both
quark and antiquark jets in \ep \ra \qqg events is \apriori more
difficult due to the greater hadronic activity, and the overall
efficiency is very low since it is proportional to the square of the
single-jet tagging efficiency.
 
It is therefore useful to consider more inclusive quantities.
Two possibilities are:
(1) {\it Semi-inclusive}: the quark jet is assumed to be identified,
and the {\it least energetic} jet in the event is taken to be the gluon
and is used to define the angle $\chi$ (eq.~{\eqchi}).
(2) {\it Fully-inclusive}:
the jets are labelled only in terms of their energies,
$x_3\le x_2\le x_1$;
the polar angle $\theta$ is then defined by the angle of the fastest
jet w.r.t. the electron beam direction\note{To \oalp this is
equivalent to the angle of the thrust axis
\ref{\refTHRUST}
{E. Farhi, Phys.\ Rev.\ Lett.\ {\bf 39} (1977) 1587.}\
w.r.t. the electron
direction.}, and the azimuthal angle $\chi$ can be defined analogously to
eq.~{\eqchi} as:
$$
\cos\chi\qu = \qu {{\vec 1}\times {\vec 3} \over{
| {\vec 1}\times {\vec 3} |}}
\cdot {{\vec 1} \times \vec{e^-} \over {
| {\vec 1} \times {\vec{e^-}}|}}
\eqnref{\eqThrchi}
$$
In both cases a similar moments analysis to that defined in Section
5 can be applied, with relations between the corresponding helicity
cross-sections and the lowest-rank moments as given in Table~1.
 
For the semi-inclusive case we have calculated the $\hat\sigma_i$
at \oalp and show the ratios
$\hat\sigmaT/(\hat\sigmaU+\hat\sigmaL)$,
$\hat\sigmaI/(\hat\sigmaU+\hat\sigmaL)$,
$\hat\sigmaP/(\hat\sigmaU+\hat\sigmaL)$ and
$\hat\sigmaA/(\hat\sigmaU+\hat\sigmaL)$
in Fig.~5b.
Whereas $\hat\sigmaP$ and $\hat\sigmaT$ are unchanged relative to the
exclusive case, $\hat\sigmaI$ and $\hat\sigmaA$, which multiply
terms proportional to $\cos\chi$ in eq.~{\eqfour}, are smaller
in magnitude because of the sometimes incorrect gluon-jet identification.
Though this implies that the parity-violating asymmetry
$\tilde A(|p|)|_{\chi}$ in eq.~{\eqAFBboth} is smaller
by a ($y_c$-dependent) factor of order 2,
it will in fact be easier to access experimentally because
the semi-inclusive case requires only one of the quark-
and antiquark-jets to be identified explicitly.
 
In the fully-inclusive case the terms $\sigmaA$ and
$\sigmaP$, which are odd under interchange of quark and antiquark jets,
cancel out; writing the cross-section in terms of thrust
{\refTHRUST}
one obtains at \oalp:
$$\eqalignno{
2\pi{\d^3\sigma \over \d(\cos\theta)\d\chi\d T}
&= {3\over8}(1+\cos^2\theta)\, {\d\sigmaU\over \d T}
+{3\over4}\sin^2\theta\, {\d\sigmaL\over \d T} \cr
&+{3\over4}\sin^2\theta\cos2\chi\, {\d\sigmaT\over \d T}
+{3\over2\sqrt{2}}\sin2\theta\cos\chi\, {\d\sigmaI\over \d T},
&(\nr)}$$
where expressions for $\d\sigma_i/\d T$ can be found in
ref.~{\refOOO}\note{Note that $\d\sigmaI/\d T$ has opposite sign
as compared with ref.~{\refOOO}, since the angle $\chi$ defined
by eq.~{\eqThrchi} is complementary to that of ref.~{\refOOO},
$\chi_{\hbox{eq.~{\eqThrchi}}}=\pi-\chi_{\hbox{ref.~{\refOOO}}}$.}.
Using the notation
$$
\tilde\sigma_i \qu \equiv \qu \int_{y_c}{\d\sigma_i\over \d T}\,\d T,
\qquad \hbox{$i=$ U, L, T, I},
$$
we have calculated the $\tilde\sigma_i$ at \oalp and show the ratios
$\tilde\sigmaT/(\tilde\sigmaU+\tilde\sigmaL)$ and
$\tilde\sigmaI/(\tilde\sigmaU+\tilde\sigmaL)$ in Fig.~5c.
Their magnitudes and dependences on $y_c$ differ relative to the
exclusive and semi-inclusive cases due to the redefinition of $\theta$
and $\chi$.
Distributions of $\cos\theta$ and $\chi$ in
this case have already been measured
and found to be in agreement with \oalp QCD calculations
\ref{\refTASSO}
{TASSO Collab., W. Braunschweig \etal,
Z. Phys.\ {\bf C47} (1990) 181.},
\ref{\refLEPone}
{
L3 Collab., B. Adeva {\it et al}, Phys. Lett. {\bf B263}
(1991) 551.
\hfill\break
DELPHI Collab., P. Abreu {\it et al},
Phys. Lett. {\bf B274} (1992) 498.
\hfill\break
SLD Collab., K. Abe \etal,
SLAC-PUB-7099 (1996); to appear in Phys.\ Rev.\ D.
}.
 
Another fully-inclusive observable is the polar-angle $\bar\theta$
of the normal to the event plane with respect to the beam direction.
The differential cross-section $\d\sigma/\d(\cos\bar\theta)$
has been calculated at \oalpsq in massless perturbative QCD
\ref{\refTHEBR}
{J.G.\ K\"orner, G.A.\ Schuler, F. Barreiro, Phys.\ Lett.\ {\bf B188}
(1987) 272.},
and has been measured at $\sqrt{s}\simeq35$~GeV {\refTASSO}
and $\sqrt{s}=91$~GeV {\refLEPone}.
The effects of final-state interactions can induce a term linear
in $\cos\bar\theta$ whose sign and magnitude depend on the electron
beam polarisation
\ref{\refDixon}
{A. Brandenburg, L. Dixon, Y. Shadmi,
Phys.\ Rev.\ {\bf D53} (1996) 1264.};
experimental limits on such a term have been set using hadronic
\z0 decays \ref{\refSLDthree}
{SLD Collab., K. Abe \etal, Phys.\ Rev.\ Lett.\ {\bf 75} (1995) 4173.}.
 
\vfill\eject
 
\noindent{\bf 7. Conclusions}
 
\vskip .7\baselineskip
 
\noindent
We have presented the orientation of \ep \ra \qqg events in terms of the
polar- ($\theta$) and azimuthal- ($\chi$) angle distributions.
These distributions have been calculated at \oalp in perturbative QCD
for massless quarks and their dependence on longitudinal electron-beam
polarisation and centre-of-mass energy has been illustrated.
We have considered the left-right forward-backward asymmetry of the
$\cos\theta$ distribution and have defined a corresponding asymmetry of
the $\chi$ distribution. Parity-violating 3-jet observables of this kind
represent a new search-ground for anomalous contributions and
have yet to be explored experimentally.
 
For the case of \epa at the \z0 resonance using longitudinally-polarised
electrons, the $\cos\theta$ asymmetry is proportional to
the QCD helicity cross-section $\hat\sigmaP$, and the $\chi$ asymmetry to
the helicity cross-section $\hat\sigmaA$, which have not yet been
measured; this should be possible, with a high-statistics data sample,
using the highly-polarised electron beam at SLC/SLD.
To lowest electroweak order the ratio of these asymmetries is
independent of electroweak couplings and the beam polarisation. These
results are valid up to \oalpsq in QCD perturbation theory.
We have calculated $\hat\sigmaP$ and $\hat\sigmaA$ at \oalp and find
their dependence on the jet resolution parameter $y_c$ to be weak. It
would be worthwhile to calculate higher-order perturbative QCD
contributions, as well as quark mass effects, before making a detailed
comparison of these predictions with data.
 
We have also defined moments of the cross-section in terms of powers of
$\cos\theta$ and $\cos\chi$, which allow the six independent helicity
contributions $\hat\sigmaU$, $\hat\sigmaL$, $\hat\sigmaT$,
$\hat\sigmaI$, $\hat\sigmaP$ and $\hat\sigmaA$ to be determined from
data.
Even the extraction of $\hat\sigmaU$, $\hat\sigmaL$, $\hat\sigmaT$ and
$\hat\sigmaI$, which does not require quark and antiquark jet
identification, represents a detailed test of QCD.

\medskip
This research has been supported by U.S.\ Department of Energy
Cooperative Agreement DE-FC02-94ER40818
and by the Research Council of Norway.
We thank Lance Dixon and Tom Rizzo for helpful discussions.
 
\vfill
 
\vfill\eject\immediate\closeout\rfile
\baselineskip=14pt\centerline{{\bf References}}\bigskip{\frenchspacing%
\catcode`\@=11\escapechar=` %
\input refs.tmp\vfill\eject}\nonfrenchspacing
 
\vskip 5truecm
{
\centerline{Table 1: Cross-section moments $\Sigma_{mn}$}
}
\vskip 12pt
{                     
\parindent=0pt
\vbox{\offinterlineskip \tabskip=0pt
\def\tablerule{\noalign{\hrule}}
\halign to \hsize{\vrule#\tabskip=6pt plus 100pt&\strut
\hfil#\hfil&\vrule#&\hfil#\hfil&\vrule#&
\hfil#\hfil&\vrule#&\hfil#\hfil&\vrule
#\tabskip=0pt\cr
\tablerule
height4pt&\omit&&&&&&&\cr
&\omit &&$n = 0     $&&$n = 1             $&&$n = 2               $&\cr
height4pt&\omit&&&&&&&\cr
\tablerule
height4pt&\omit&&&&&&&\cr
&\omit\hfil$m = 0             $&&
\omit $(\hat\sigmaU+\hat\sigmaL)\,h_f^{(1)}(s)$ &&
\omit $-{3\pi\over4\sqrt2}\hat\sigmaA\,h_f^{(2)}(s)$ &&
\omit $(\half\hat\sigmaU+\half\hat\sigmaL+\fourth\hat\sigmaT) \,
h_f^{(1)}(s)$ &\cr
height4pt&\omit&&&&&&&\cr
\tablerule
height4pt&\omit&&&&&&&\cr
&\omit\hfil$m = 1                       $&&
\omit $\half\hat\sigmaP\,h_f^{(2)}(s)$ &&
\omit ${3\pi\over16\sqrt2}\hat\sigmaI\,h_f^{(1)}(s)$ &&
\omit $\fourth\hat\sigmaP\,h_f^{(2)}(s)$ &\cr
height4pt&\omit&&&&&&&\cr
\tablerule
height4pt&\omit&&&&&&&\cr
&\omit\hfil$m = 2                         $&&
\omit $({2\over5}\hat\sigmaU+{1\over5}\hat\sigmaL) \,
h_f^{(1)}(s)$ &&
\omit $-{3\pi\over16\sqrt2}\hat\sigmaA\,h_f^{(2)}(s)$ &&
\omit $({1\over5}\hat\sigmaU+{1\over10}\hat\sigmaL
+{1\over20}\hat\sigmaT) \, h_f^{(1)}(s)$ &\cr
height4pt&\omit&&&&&&&\cr
\tablerule}}
}                
 
\vfill\eject
 
 
\noindent{\bf Figure Captions}
 
\vskip .5truecm
 
\noindent
Fig.~1. Tree-level Feynman
diagrams for three-jet production in \ep   annihilation.
 
\vskip .5truecm
 
\noindent
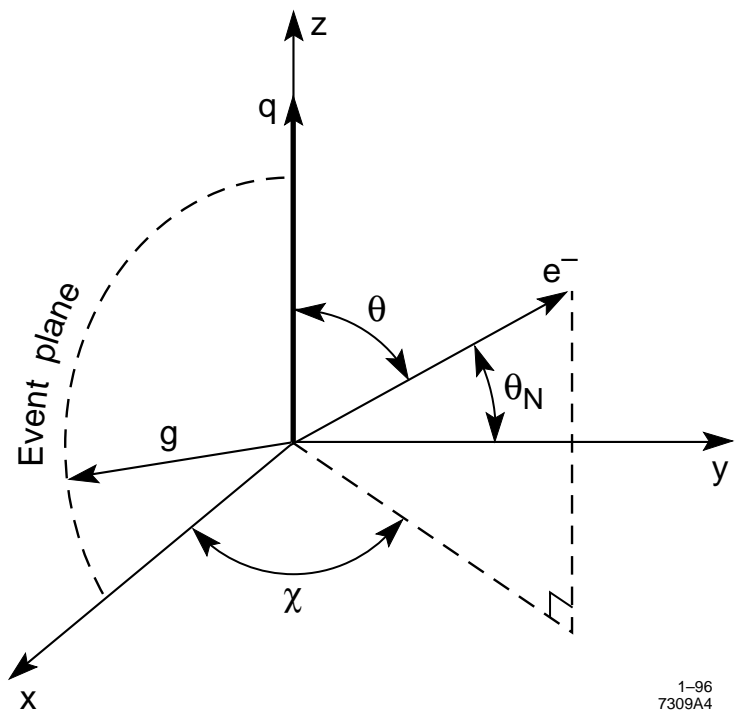
Fig.~2: Definition of the angles $\theta$ and $\chi$.
 
\vskip .5truecm
 
\noindent
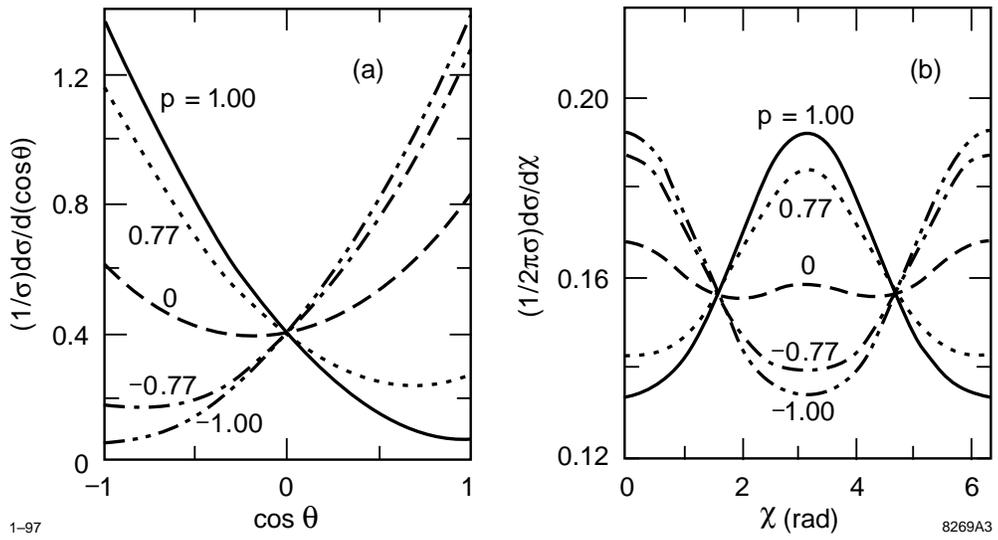
Fig.~3: Angular orientation of
the event plane for down-type quarks and $y_c=0.02$.
Distribution of (a) $\cos\theta$ and (b) $\chi$
at $\sqrt{s}=M_Z$, for 5 values of $p$.
 
\vskip .5truecm
 
\noindent
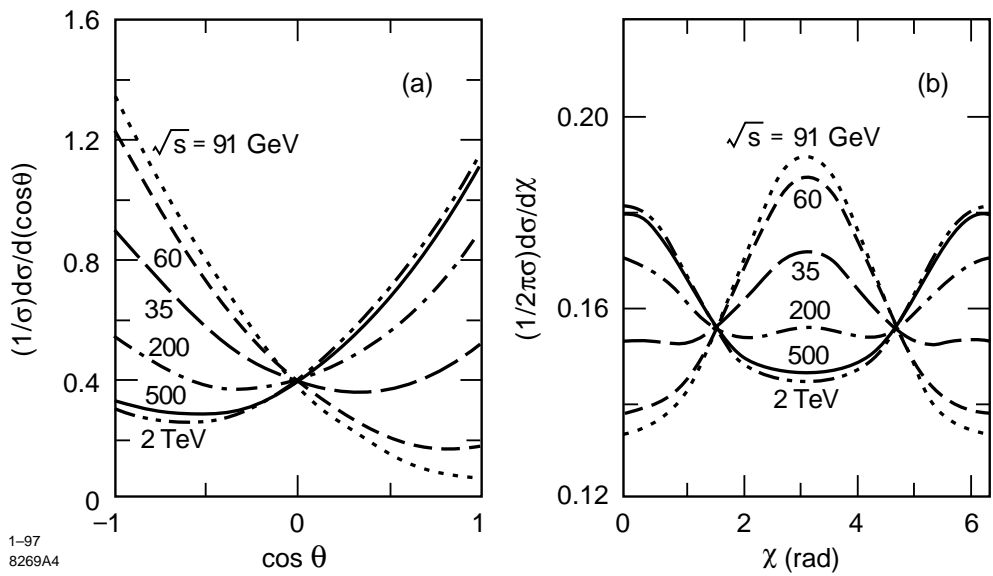
Fig.~4: Angular orientation of
the event plane for down-type quarks and $y_c=0.02$.
Distribution of (a) $\cos\theta$ and (b) $\chi$
for $p=+1$ at 6 values of $\sqrt{s}$.
 
\vskip .5truecm
 
\noindent
Fig.~5: Helicity cross-section ratios
(see text) as functions of $y_c$;
(a) exclusive, (b) semi-inclusive and (c) fully-inclusive cases.
For the sake of clarity, the ratios are multiplied by
a factor of 5 where indicated.


 
 
 
 
 
 
\bye